\documentclass[aps,prl,preprint,groupedaddress]{revtex4}
\begin{document}

%\preprint{}

\title{Born Rule and  Finkelstein-Hartle Frequency Operator Revisited}


\author{Patrick Das Gupta}
\email[]{patrick@srb.org.in}
%\author{Debapriya Chaudhuri}
%\thanks{}
\affiliation{Department of Physics and Astrophysics, University of Delhi, Delhi - 110 007 (India)}


\date{\today}

\begin{abstract}
%With the gleaning of graphene, stable two dimensional crystals have successfully materialized  in the laboratories.
Character of observables in classical physics and quantum theory is reflected upon. Born rule in the context of measurement being an interaction between two quantum systems is discussed.
A pedagogical introduction to Finkelstein-Hartle frequency operator is presented.
\end{abstract}

%\keywords{}

\maketitle

%\begin{document}       

%\author{Patrick Das Gupta}
%\begin{abstract}
%With the gleaning of graphene, stable two dimensional crystals have successfully materialized  in the laboratories.

%\end{abstract}
\section {Introduction}         
According to quantum theory,  the physical state  of a system is represented by a normalized state-vector $|\psi>$ lying in a Hilbert space corresponding 
to the system. 
In the absence of any measurement performed on the system, as  time rolls out, this state-vector $|\psi(t)>$ is found to evolve unitarily,
 ordained by the 
Hamiltonian of the system. Such an evolution, led by  Schrodinger equation, is causal and deterministic. Physical quantities associated with a system that can be measured are
termed as observables, and are represented by self-adjoint operators defined on the Hilbert space. Hamiltonian is a special linear operator corresponding to the energy observable of the
system that is responsible for  dynamics.

 There is a subtle difference between the roles played by
observables in classical physics and those in quantum mechanics (QM), other than the measurement limitation imposed by Heisenberg uncertainty principle
 on a pair of non-commuting observables in QM. 
In the realm of classical theories, the physical state of a system is completely specified by a set of independent observables of the system. There is no fundamental distinction between 
the physical state and a complete list of measured values of observables. Generically, the operational definition of a physical 
  observable associated with a classical particle involves directly its state, and the definition itself suggests a method by which one can measure
 it. 

 As an illustration, if one were to measure the velocity of a particle at time $t$, one could just measure the positions at  times $t$ and 
  $t+\Delta t$, given an infinitesmally tiny $\Delta t$ (note that the position variable is associated with the state of the particle). 
The ratio of  difference between the two positions to $\Delta t$ is  the measured velocity of the particle,
since that is how velocity is defined in the first place. As an aside, this is one of the primary reasons why velocity is ill-defined in QM, because a 
 precise measurement of position of the particle at an instant renders its momentum completely uncertain, making the subsequent
 position measurement impossible (the other reason being that special relativity along with uncertainty principle imply that localization of a particle to a region of size less than its
 Compton wavelength necessitates creation of other particles, thereby raising doubts about single particle QM and position being a fundamental observable).

On the other hand, in quantum physics, an observable  has an abstract representation in terms of a hermitian operator defined independently of the
state. This fundamental separation between states and observables is a quantum feature. Now, the  form of a hermitian operator by itself does not suggest a way
 to measure it. Instead, one frequently invokes classical concepts and physical intuition to arrange for a  suitable interaction between the system and the apparatus such that 
the  observable 
(or, a related operator like its canonical conjugate, etc.) 
 gets coupled with a `pointer variable' associated with the measuring apparatus, for the purpose of 
measurement.  To cite an example, the expression for  the all too familiar 
spin angular momentum operator of an electron is given by,

$$\hat {\vec S}= \frac {1} {2} \hbar \vec \sigma ,$$

 with $\sigma_i , i=1,2,3$  being the Pauli matrices. Evidently, $\hat {\vec S}$ does not involve the general state of the electron, and its mathematical form (or, the commutator brackets
 of components of the spin)  does 
 not give us any clue  as to how to go about
 measuring it.

 For the measurement of  electron spin in a Stern-Gerlach like contraption,
 one relies on the classical picture that a spinning charge particle has a magnetic moment  $\vec \mu $, so that in the presence of an 
inhomogeneous  magnetic field $\vec B (\vec r) $, the
interaction energy  -$\vec \mu . \vec B $  leads to a coupling between the spin degree of freedom and the position (`pointer variable') of the electron. Then, assuming that the 
magnetic moment operator is related to the spin,
$$\hat {\vec \mu} \propto \hat {\vec S} ,$$
one obtains an interaction Hamiltonian,
$$\hat H_{int}= - \hat {\vec \mu}  . \vec B \propto - \hat {\vec S} . \vec B$$
If the initial spin state is orthogonal to $\vec B$, time evolution due to the above interaction Hamiltonian causes, by virtue of Schrodinger equation, an entanglement between spin
 and position degrees such that the state of the electron is described by  a superposition of correlated electron's spin and position states. 

However, when a device like a photographic plate is deployed to measure the position
of an electron, one observes a definite position correlated with a spin state, and not a superposition of correlated states predicted by the unitary evolution (this is the central mystery of QM). 
The spin state, thereafter, is inferred from the 
 observed position.
 Attempts to understand how the entangled state  
 breaks into two distinct branches corresponding to spin up and spin down have kept the QM community active for the last eight decades or so (see for example [1]).

 In general, the outcome of a measurement is always one of the eigenvalues $a_i, i= 1,2,...$ of the self-adjoint operator $\hat A $ 
 that represents the measured observable. But which one? It is at this point that 
the apparition of indeterminism makes its presence felt. Normally, one cannot predict with certainty the  outcome of a measurement. 

 More importantly,  Max Born had discovered  that
 the probability  of finding a particular eigenvalue $a_i$ to be the 
outcome  at time $t$ is given by $|<i|\psi (t)>|^2$ and, according to standard QM, it is associated with a `collapse' of the state-vector $|\psi(t)>$ to
 $|i>$, an eigenstate of $\hat A$ with eigenvalue $a_i$. The probabilities of the potential outcomes can definitely be predicted with
certainty.  How is the Born rule verified in practice?

The standard probabilistic nature of QM emerges from measurements conducted on an ensemble consisting of a large number N of identical systems that are isolated from each other and
 placed in identical
surroundings, with
 each being described by an identical state-vector  $|\psi(t_0)>$ at time $t_0$. Thereafter, the  state of each system evolves unitarily in an identical manner, governed by the Schrodinger
 equation.
 When one performs at some time $t$ measurements of  an observable, say $\hat A$, on each system belonging to the ensemble, one finds a random distribution of measured values with each
 outcome being one of the eigenvalues $a_i, i=1,2,..$, corresponding to a measurement. Then, it turns out that the frequency of occurrence of the eigenvalue $a_i$ is $|<i|\psi (t)>|^2$, in
the limit N tending to a very large number.

 Although from a practical point of view quantum mechanics is very successful, issues related to measurements have continued to perplex one since the inception of the theory.
 Any measurement after all
is a physical interaction between the system and the apparatus (which too is governed by laws of QM)
 involving a definite interaction Hamiltonian, so that one expects the combined system to evolve with time  according to  Schrodinger 
equation in a deterministic manner. In that case, why are outcomes of measurement acausal and unpredictable? Stated differently, from where does  Born rule spring from? 

In the framework 
of Many-Worlds Interpretation (MWI)  of QM, the state of the combined system continues to be an entanglement of system-apparatus-observer correlated states as ordained by 
 unitary evolution, and there never occurs a wavefunction collapse  in this picture [2]. Can the observed indeterminism in QM be explained by MWI?  There
have been humongous amount of research work on the issue of obtaining Born rule without invoking probabilistic `collapse of the state-vector to an eigenstate' in MWI, as well as 
other approaches like, for instance, decoherence framework (see [1-5] and the references therein).

In this lecture, we ask a different question. Can we have Born probabilities
 as eigenvalues of a frequency observable that is  defined over an ensemble?  It is in this context that
 Hartle's paper assumes significance since it had attempted to deduce  Born probabilities from frequencies of occurrences of eigenvalues in measurements in a framework 
  describing
 an ensemble of large number of identical systems as an individual quantum system over which a frequency operator is defined [6].  In fact, the form of the frequency observable was introduced 
even earlier by Finkelstein [7], and therefore is referred to  as the Finkelstein-Hartle frequency operator [8].  In the following section, we present a pedagogical description of
 Hartle's approach.

\section {Finkelstein-Hartle frequency operator}

  The basic idea is to devise an observable   whose eigenvalues are the Born probabilities (without invoking the Born rule). If the collection of measurements over an ensemble is viewed as a
 grand measurement of this frequency observable, then it follows from  laws of QM that only its eigenvalues are the outcomes. Suppose the eigenvalues turn out to be $|<i|\psi (t)>|^2$, then
one has moved one step forward in understanding the origin of indeterminism in QM.

We begin with a quantum system described by a Hilbert space $H$ and an observable $\hat A $ defined on it. For simplicity, we assume that $\hat A$  has  a discrete spectrum of eigenvalues so that,
$$\hat A |i\rangle = a_i |i\rangle,  i=1,2,3,.....\eqno(1a)$$
 with eigenstates $| i\rangle $ forming a complete 
 orthonormal basis, satisfying the inner product  orthonormality,
$$ (| i\rangle, | j\rangle) = \langle i | j\rangle =\delta _{ij}\eqno(1b)$$
and completeness,
$$\sum_{i=1}^\infty |i\rangle \, \langle i | = 1 \  ,\eqno(1c)$$ 
where $a_i $ are the eigenvalues that are observed when $\hat A$ is measured.

A physical state of the system is described  by  $|\psi   \rangle$, which is an element of $H$,  and can be linearly expanded in terms of the orthonormal basis  vectors $\lbrace |i\rangle, \ i = 1, 2\dots  \rbrace$,
$$|\psi  \rangle=\sum_{i=1}^\infty  {c_i |i\rangle}\ ,\eqno(2a)$$
with,
$$\langle \psi | \psi \rangle = 1 \ ,\eqno(2b)$$
and,
$$c_i=  \langle i | \psi \rangle \ ,\eqno(2c)$$
which follows from eqs.(1b) and (2a).

Since, the probabilistic character of QM (concerning outcomes of measurements) have been tested by making use of large number of identically prepared systems, we need to formulate 
the measurement problem
accordingly.  An ensemble of  N identical systems in QM is represented by the Hilbert space formed out of the tensor product of individual spaces $H \times  H \times \dots \times H \equiv H^N$.

For convenience, we label the systems belonging to the ensemble using the index $\alpha $ with $\alpha=1,2,...,N$.
If each system is specified by the state-vector $|\psi \rangle$,  the physical state describing the ensemble is then represented by an element of $H^N$ given 
by the
direct product of $|\psi \rangle$s,
$$|(\psi)^N\rangle \equiv |\psi \rangle_1 |\psi \rangle_2 \dots |\psi \rangle_N , \eqno(3a)$$
where the  subscript on $|\psi \rangle $ indicates the system to which the state-vector  corresponds.

In the limit $N \rightarrow \infty$, the ensemble state-vector tends to,
$$|(\psi)^\infty\rangle \equiv |\psi \rangle_1 |\psi \rangle_2 \dots |\psi \rangle_N |\psi\rangle_{N+1}\dots ,\eqno(3b)$$
assuming that the limit is well-defined (one has to be careful with such limits, as most of the peculiarities which we  come across later, stem from such infinities
  and associated measures [8]).

 We use a notation in which, for the $\alpha^{th}$ system of the ensemble,  $|i_\alpha \rangle $ denotes the eigenstate of $\hat A$ with 
eigenvalue $a_{i_\alpha }$ , 
so that  $\lbrace |i_\alpha \rangle ,  i_\alpha=1,2,\dots \rbrace $ is an orthonormal basis  corresponding to the $\alpha^{th}$ system  (see eqs.(1a)-(1c)). The Hilbert space $H^N$ is therefore
 spanned 
by the direct product of orthonormal vectors $\lbrace |i_1 \rangle |i_2 \rangle \dots |i_N \rangle , \ \ i_1, i_2, \dots = 1,2,\dots \rbrace$.

These direct product of eigenstates and their dual can be used to construct a frequency operator  $F^j_N$ for the eigenvalue $a_j$ of $\hat A$ as follows,
$$\hat F^j_N \equiv \sum_{i_1, i_2, \dots ,i_N}{f_j |i_1 \rangle |i_2 \rangle \dots |i_N \rangle
 \langle i_N | \langle i_{N-1}| \dots \langle i_2| \langle i_1| } \ ,\eqno(4a)$$
where,
$$f_j \equiv {1\over{N}} \sum_{\alpha=1}^N {\delta _{j i_\alpha}}\eqno(4b)$$
is clearly the frequency of $i_\alpha$ being equal to $j$ in the set $\lbrace i_1, i_2, \dots , i_N \rbrace$.

It is easy to see that $|i'_1 \rangle |i'_2 \rangle \dots |i'_N \rangle $ is an eigenstate of $F^j_N$ corresponding to the eigenvalue being the
frequency of $i'_\alpha$ equal to $j$ for $\alpha=1,2,\dots,N$
in $\lbrace i'_1, i'_2, \dots, i'_N \rbrace $, since from eqs.(4a-b) and the orthonormality of $\lbrace |i_\alpha \rangle,  i_\alpha = 1,2, \dots \rbrace $ (eq.(1b)) we get,
$$\hat F^j_N |i'_1 \rangle |i'_2 \rangle \dots |i'_N \rangle = \sum_{i_1, i_2, \dots ,i_N}{{1\over{N}} \sum_{\alpha=1}^N {\delta _{j i_\alpha}} |i_1 \rangle |i_2 \rangle \dots |i_N \rangle \delta_{i_1 i'_1} \delta_{i_2 i'_2} \dots \delta_{i_N i'_N}}\eqno(5a)$$
$$= {1\over{N}} \sum_{\alpha=1}^N {\delta _{j i'_\alpha}} \bigg (|i'_1 \rangle |i'_2 \rangle \dots |i'_N  \rangle \bigg ) \ \ ,\eqno(5b)$$
 thus, vindicating that $F^j_N $ indeed is a frequency operator.
 
 For later purposes, it is useful to express the frequency operator as,
 %\begin{eqnarray*}
$$   \hat F^j_N = {1\over{N}} \sum_{i_1, i_2, \dots ,i_N}{|i_1 \rangle |i_2 \rangle \dots |i_N \rangle \bigg (\delta_{j i_1} + \delta_{j i_2}  + \dots + \delta_{j i_N} \bigg ) \langle i_N | \langle i_{N-1}| \dots \langle i_2| \langle i_1| } $$
 $$= {1\over{N}} \bigg \lbrace |j \rangle _{1  1}\langle j|  \sum_{i_2, i_3, \dots ,i_N}{ |i_2 \rangle \langle i_2 |  |i_3 \rangle \langle i_3 | \dots   |i_N\rangle \langle i_N| } + $$
  $$+  |j \rangle _{2   2}\langle j|  \sum_{i_1, i_3, \dots ,i_N}{ |i_1 \rangle \langle i_1 |  |i_3 \rangle \langle i_3| \dots   |i_N\rangle \langle i_N| }+ \dots +  $$
 $$+  |j \rangle _{N  N}\langle j|  \sum_{i_1, i_2, \dots ,i_{N-1}}{ |i_1 \rangle \langle i_1|  |i_2 \rangle \langle i_2 | \dots   |i_{N-1}\rangle \langle i_{N-1}| }\bigg \rbrace $$
$$={1\over{N}} \bigg \lbrace |j \rangle _{1  1}\langle j|   +
   |j \rangle _{2   2}\langle j|  +\dots 
 +  |j \rangle _{N  N}\langle j|  \bigg \rbrace \eqno(6)$$
% \end{eqnarray*}
 The last step follows from eq.(1c). In eq.(6), the ket $|j \rangle _\alpha $ and its dual represent the eigenstate $|j \rangle $ and the eigenbra, respectively, 
corresponding to the eigenvalue $a_j$ for the $\alpha^{th}$ system.

The operation of $\hat F^j_N$ on $|(\psi)^\infty \rangle $ is defined by,
 $$\hat F^j_N |(\psi)^\infty \rangle \equiv (\hat F^j_N |(\psi)^N \rangle ) |\psi\rangle_{N+1}|\psi\rangle_{N+2}\dots  \eqno(7a)$$
 Using eq.(6) and eq.(2c), we obtain,
 $$\hat F^j_N |(\psi)^N \rangle ={{c_j}\over{N}} \bigg \lbrace |j \rangle _1  |\psi \rangle_2 |\psi \rangle_3\dots |\psi \rangle_N + 
     |\psi \rangle_1 |j \rangle_2  |\psi \rangle _3 \dots |\psi \rangle_N + $$
 $$ + \dots +   |\psi \rangle_1 |\psi \rangle_2 \dots |\psi \rangle_{N -1}  |j \rangle _N \bigg \rbrace \ , \eqno(7b)$$
 so that, from eq.(7a), we have,
  $$\hat F^j_N |(\psi)^\infty \rangle ={{c_j}\over{N}} \bigg \lbrace |j \rangle _1  |\psi \rangle_2 |\psi \rangle_3\dots |\psi \rangle_N +
     |\psi \rangle_1 |j \rangle_2  |\psi \rangle _3 \dots |\psi \rangle_N + $$
 $$ + \dots +   |\psi \rangle_1 |\psi \rangle_2 \dots |\psi \rangle_{N -1}  |j \rangle _N \bigg \rbrace  |\psi\rangle_{N+1}|\psi\rangle_{N+2}\dots  \eqno(7c)$$
Following Hartle [6], we may ask how close is the state-vector $ \hat F^j_N |(\psi)^\infty \rangle $ to $|c_j|^2   |(\psi)^\infty \rangle $ in the limit  $N \rightarrow \infty $? Now, to address
this question, we may use the norm that is induced by the inner product, in order to obtain,
 
$$||  \hat F^j_N |(\psi)^\infty \rangle  -     |c_j|^2   |(\psi)^\infty \rangle   ||^2 \equiv  \bigg (  \hat F^j_N |(\psi)^\infty \rangle  -     |c_j|^2   |(\psi)^\infty \rangle ,  \hat F^j_N |(\psi)^\infty \rangle  -     |c_j|^2   |(\psi)^\infty \rangle \bigg )\eqno(8a)$$
 $$= \bigg ( \hat F^j_N |(\psi)^\infty \rangle ,  \hat F^j_N |(\psi)^\infty \rangle \bigg) - 2 |c_j|^2 \bigg ( |(\psi)^\infty \rangle ,  \hat F^j_N |(\psi)^\infty \rangle \bigg ) + |c_j|^4  \eqno (8b)$$
 The last term in eq.(8b) arises from the fact that  $\langle (\psi)^\infty   |(\psi)^\infty \rangle = 1$ because of eqs. (2b) and (3b). Again, eqs.(2b), (2c), (3b) and (7c) lead to the inner product,
 $$\bigg ( |(\psi)^\infty \rangle ,  \hat F^j_N |(\psi)^\infty \rangle \bigg ) = {{c_j}\over{N}} \bigg  \lbrace  {}_1\langle \psi | j\rangle_1 + {}_2 \langle \psi | j\rangle_2 + \dots + {}_N  \langle \psi | j\rangle_N \bigg \rbrace 
 = |c_j|^2 \eqno(8c)$$
 
 From eqs.(7c) and (2b), we get,
 $$\bigg ( \hat F^j_N |(\psi)^\infty \rangle ,  \hat F^j_N |(\psi)^\infty \rangle \bigg) = {{|c_j|^2}\over{N^2}}  \bigg \lbrace  {}_1 \langle j  | {}_2 \langle \psi | {}_3 \langle \psi | \dots  {}_N \langle \psi | + $$
  $$ +  {}_1 \langle \psi | {}_2 \langle j | {}_3 \langle \psi | \dots  {}_N \langle \psi | + \dots + {}_1 \langle \psi | {}_2 \langle \psi | {}_3 \langle \psi | \dots  {}_{N-1} \langle \psi | {}_N \langle j  | \bigg \rbrace  \times$$
  $$ \times \bigg \lbrace |j \rangle _1  |\psi \rangle_2 |\psi \rangle_3\dots |\psi \rangle_N +
     |\psi \rangle_1 |j \rangle_2  |\psi \rangle _3 \dots |\psi \rangle_N + 
  \dots +   |\psi \rangle_1 |\psi \rangle_2 \dots |\psi \rangle_{N -1}  |j \rangle _N \bigg \rbrace \eqno(9a)$$
 $$={{|c_j|^2}\over{N^2}} \lbrace N  + N(N-1) |c_j|^2 \rbrace \eqno(9b)$$
 Hence, employing eqs.(8c) and (9b) in eq.(8b), give rise to,
 $$||  \hat F^j_N |(\psi)^\infty \rangle  -     |c_j|^2   |(\psi)^\infty \rangle   ||^2  = {{|c_j|^2}\over{N}} \lbrace 1- |c_j|^2 \rbrace \eqno(9c)$$
 From eq.(9c), it is evident that  as  $N \rightarrow \infty $,  we have $||  \hat F^j_N |(\psi)^\infty \rangle  -     |c_j|^2   |(\psi)^\infty \rangle   ||^2  \rightarrow 0$. 
 
 This is a remarkable result in the sense that no matter what $|\psi \rangle $ is, for every eigenvalue $a_j$ of the observable $\hat A $, the distance between
 the state $\hat F^j_N |(\psi)^\infty \rangle $ and the Born probability times 
  $ |(\psi)^\infty \rangle $ can be made arbitrarily small by considering sufficiently large ensemble. But this by no means implies that as  $N \rightarrow \infty $, the state-vector
  $ \hat F^j_N |(\psi)^\infty \rangle $  $\rightarrow  |c_j|^2   |(\psi)^\infty \rangle $.  

In fact, it can be  demonstrated that the vanishing of the left hand side of  eq.(9c) does not entail that $\hat F^j_N |(\psi)^\infty \rangle $ =
    $|c_j|^2   |(\psi)^\infty \rangle $ as  $N \rightarrow \infty $ [8,9]. As much  is hinted by the expression in the right hand side of eq.(7b). 

The hope articulated in the beginning of this section of  obtaining  Born probabilities  $ |c_j|^2 $, $j=1,2, \dots $ as eigenvalues of the frequency operator remains unfulfilled in this
 approach (For a detailed
discussion on this issue please refer to the papers by Squires [9], Caves and Schack [8] as well as N. D. Hari Dass' lecture in this meeting).

\section {Summary}

The preceding section draws our attention to some very interesting points. From eqs.(8c) and (9c), we find that the expectation value of the frequency operator and 
its uncertainty corresponding to
 the state $|(\psi)^\infty \rangle$ are given by,
$$\langle F^j_N \rangle = |c_j|^2 \eqno(10a)$$
and,
$$\Delta F^j_N =\sqrt {\frac {|c_j|^2 - |c_j|^4} {N}}, \eqno(10b)$$
respectively. Although in the limit N tending to infinity, $\Delta F^j_N $ vanishes,  the state-vector  $|(\psi)^\infty \rangle $ describing the ensemble does not become an eigenstate of the
frequency operator with the Born probability $|c_j|^2$ as the eigenvalue (For a thorough critical analysis, please see the paper by Caves and Schack [8]). 
 
The frequency operator approach is silent about how to measure the corresponding observable. If one employs the obvious method of measuring $\hat A$ for every system in the
 ensemble, to find the frequency of
 occurrence of
 an eigenvalue, then this approach
 does not throw much light on the measurement problem as to whether $|\psi>$ collapses to one of the  eigenstates in  individual measurements.
The enigma of  Born rule continues to be wrapped in a riddle inside a mystery!
    
\section {Acknowledgements}
It is a pleasure to thank Professor N.D. Hari Dass and Professor R. Srikanth for providing a stimulating atmosphere for debates and discussions during the meeting, and also for their generous
 hospitality at PPISR, Bangalore.

 %\pagebreak
 {\bf References}
 
 [1] Wheeler, J. A. and Zurek, W. H. (eds), 1983, "Quantum Theory and Measurement", Princeton: Princeton University Press, and the references therein.

 [2] Everett, H., 1957, "Relative State Formulation of Quantum Mechanics", Reviews of Modern Physics, 29: 454-462
 
 [3] DeWitt, B. S., and  Graham, N. (eds.), 1973, "The Many-Worlds Interpretation of Quantum Mechanics", Princeton: Princeton University Press.

 [4] Zurek W. H., "Probabilities from entanglement, Born's rule from envariance", 2005, Phys. Rev. A 71: 052105 

 [5] Buniy, R. V., Hsu, S.D.H. and Zee, A., "Discreteness and the origin of probability in quantum mechanics", 2006, Phys.Lett. B640:  219-223
 
 [6] Hartle, J. B., 1968, "Quantum Mechanics of Individual Systems", Am. Jour. Phys., 36: 704-712

 [7] Finkelstein, D., 1963, "Logic of Quantum physics", Transactions of the New York Academy of Sciences, 25: 621- 637 

 [8] Caves, C. M. and Schack, R., "Properties of the Frequency Operator do not imply the Quantum Probability postulate", 2005, Ann. Phys. 315: 123 - 146 
 
 [9] Squires, E. J., 1990, “On an alleged 'proof' of the quantum probability law,”, Phys.Lett.A, 145, 67

\end{document}